# A Plural Decentralized Identity Frontier: Abstraction v. Composability Tradeoffs in Web3[*]

**Shrey Jain**[†]

University of Toronto
shreyjaineth@gmail.com

**Leon Erichsen**

Gitcoin
leon@gitcoin.co

**Glen Weyl**

Microsoft, RadicalXChange
glen@radicalxchange.org

Designing a system that is both abstractable—or generalizable—and composable in unique settings is difficult. This canonical generalization problem is often wrestled amongst designers of AI systems and standards committees. If you are building in web3 today, you may be asking questions like:

1) "Will my solution work across *all* blockchains?"

2) " Is this *the* blockchain I want to be building on?"

3) "The blockchain is 'immutable,' is my solution *permanent*?"

In this article, we explore the tension between abstraction and composability in web3 today, specifically within identity solutions, and argue that **the current standard DID v1.0 is sufficiently under-specified, allowing for many methods and instantiations,[1] including blockchain-based certificates.**

We view today's experiments in web3 identity as **additive and complementary**, and argue that often-cited differences are of degree and more in form, less in substance. By way of illustration, we compare decentralized naming services and blockchain-based identity certificates such as soulbound tokens (SBTs) to decentralized identifiers (DIDs) and verifiable credentials (VCs). Both paradigms—to the extent they can be meaningfully

[1] For an analysis of well referenced DID methods and their properties based on the categorization specified in the W3C recommendation rubric, see https://eprint.iacr.org/2021/1087.pdf.

[†] Shrey Jain is the corresponding and lead author. Any questions should be directed to his email for prompt response.

[*] This would not be possible without the contributions and reviews from amazing people. We will only name a few: Yorke E Rhodes III, Vitalik Buterin, Enrico Bottazzi, Puja Ohlhaver, Phil Daian, Andrew Miller, JB Rubinovitz, Paula Berman, Joel Thorstensson, Evin McMullen, and Jonny Howle. There was productive discourse amongst this group of reviewers, but the article does not necessarily represent the views of everyone listed here. All errors and views are our own and do not represent the views of our employers, affiliated institutions, or reviewers.

differentiated[2]—share similar potential as well as challenges. Specifically, we refer to fears about non-consensual verification—or "scarlet letters"— and show DID method iterations are not immune by issuing an innocuous public "scarlet letter" to a DID's associated public address for anyone to see. Moreover, we argue that because SBTs are unspecified, one could characterize SBTs as an iteration—or extension—of VCs that additionally aspire to achieve composability with web3 smart contracts for correct execution of code, privacy, coercion resistance, and censorship resistance. We offer research paths for how VCs can also achieve these properties. We do not comment on cost, scalability, transferability, or common knowledge as they have been previously reviewed.

Our aim is to not elevate one paradigm over the other. Rather, our hope is to encourage collaboration between developer communities experimenting with non-transferable tokens and VCs to innovate **better, socially programmable identity primitives across contexts and use cases.**

## Standards Paradox

The generalization problem in web3 can be illustrated by Image 1 below. The challenge is to ensure that solutions work across as many platforms as possible while maintaining the composability within each ecosystem.

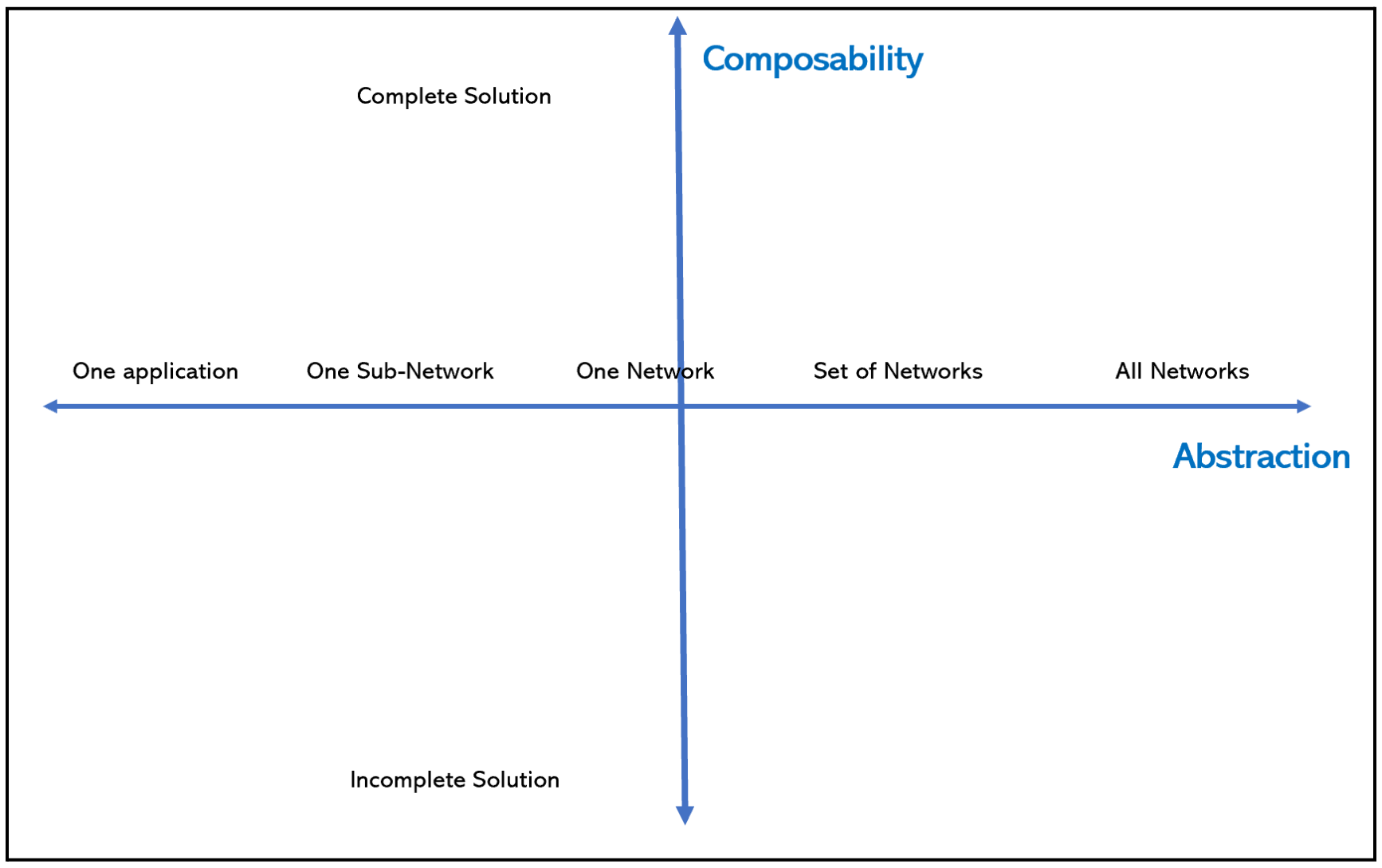

**Image 1.** Composability and abstraction tradeoff matrix.

[2] The paper Decentralized Society: Finding Web3's Soul sketched SBTs from the perspective of social incentive design but did not provide a specification. While the paper posited a first use case of tokens with the properties of 1) non-transferability 2) initial publicity 3) possibly revocability by the issuer, the authors made clear these properties were merely a bootstrapping stepping stone for initial use cases on credentials and affiliations that many—including critics— make public on platforms like Twitter and LinkedIn today. Indeed, the authors envision the degree of privacy/publicity and revocability to be encoded by the social group issuing the SBT through socially programmable privacy. The authors also emphasize community wallet recovery as a necessary but yet-to-be innovated mechanism to enforce non-transferability of SBTs. VCs share the same potential for social programmable privacy, and DIDs for community wallet recovery.

One way to think about the problem is by way of analogy to a couple's vacation compared to a larger group vacation, as sketched by Michael Mignano in his recent 'Standards Innovation Paradox' blog. Mignano artfully explains the innovation ceiling you hit early by adopting generalizable standards too soon.

> Imagine you and your significant other are alone together on a vacation for two weeks in a country you've never visited before. Because it's just the two of you, you can do anything you want on that trip without putting much thought into it. Want to cancel tonight's dinner reservation and go to a concert instead? You can. Want to skip tomorrow's museum visit and instead rent a car to go to a different city? You can.
>
> Now, imagine that same trip, but instead of it just being the two of you, your kids, your parents, your in-laws, three friends, your brother, his partner, and their four kids all tag along, too. It's a completely different trip, right? In this version of the trip, everything has to be planned meticulously. And if you decide you want to make changes to the itinerary, you have to get everyone to agree, which is nearly impossible. What you end up with is a great time spent with family and friends you haven't seen in a while, but a consensus-driven trip that is far less interesting and unique. That's what it's like building products based on standards that have achieved scale and widespread adoption.
>
> Anytime a team wants to do something exciting and new that exceeds the limitations of the standard, they have to get every stakeholder (or at least enough to reach a critical mass of adoption) who has adopted that standard to also adopt the change, otherwise the change is useless. And if you plow ahead with the change anyway and break the standard, then you lose the benefits of the standard. This is hard enough with a bunch of friends and family on a vacation, but just imagine trying to do it with a variety of companies, big and small, all with different and potentially competing interests and priorities. This is the paradox of building with standards.

Who is web3? The traveling couple or group? Let's consider both sides.

On the one hand, one could argue that the traveling couple are analogous to the novel, rapid, pivoting experiments running in small web3 communities, and the group vacation is what web3 becomes with adoption of 'standards.' Given web3's nascency, it is impossible to know in advance if these standards would satisfy the changing needs and preferences of the couple. In particular, given that the web3 ecosystem has strong values against surveillance and censorship, continued experimentation on approaches is worthwhile, especially since web3 contains minimal sensitive identity data (healthcare, government, insurance). Settling for a local maxima identity standard out of convenience risks adopting a solution that stunts and limits web3's innovation potential—possibly by orders of magnitude. Image 2 from Mignano's blog shows the innovation ceiling hit with prematurely adopting standards.

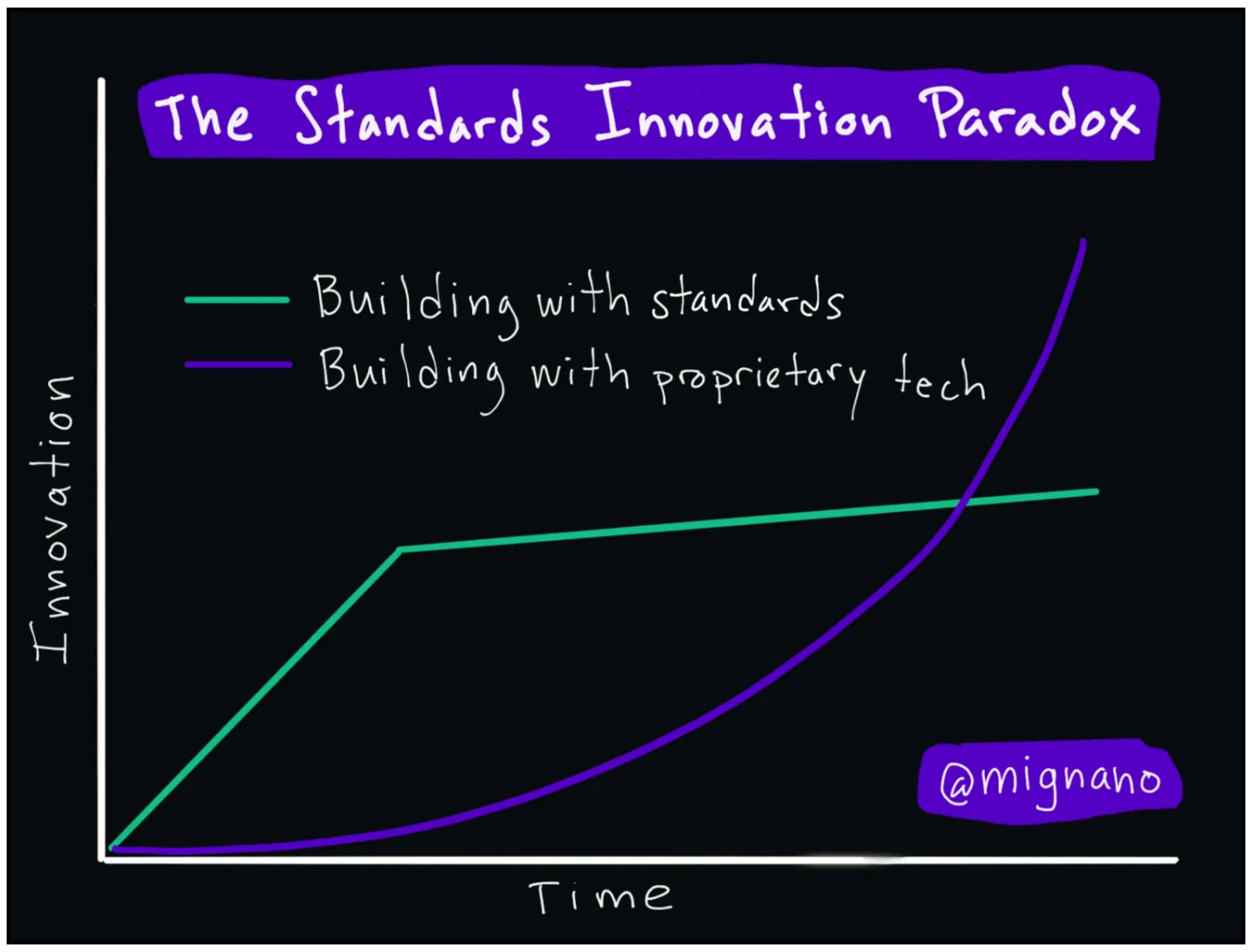

**Image 2.** From 'The Standards Innovation Paradox.'

On the other hand, an argument favoring adopting standards early on is that they can yield positive externalities. The more parties that adopt a standard, the greater coordination and network effect. As Mignano notes, standards create a "common language for how products can interact with other products, eliminating the need to build each component within a market or re-define how systems communicate with each other." And it is vital that we eventually build some consensus on how to approach a problem. For example, thanks to TCP / IP (1970s), SMTP (1980s), RSS (1999), and FHIR (2011), we can benefit from a rich ecosystem on the internet of mail, podcasts, and information.

We argue a third way. As noted by the initial draft dates of each of these standards, it can take years of iteration of an initial specification before becoming a widely adopted standard. For example, the FHIR specification—initially drafted in 2011—is only now reaching global adoption. We argue identity standards are sufficiently under-specified today, enabling innovation and expansion across various architecturals substrates; we are still early.

## Standards in Web3 Identity Today

The internet was not designed with an identity layer. For years, diverse and passionate communities sought solutions to address this problem. On July 19th, 2022, the World Wide Web Consortium (W3C) announced that DID v1.0—the DID specification initially proposed by the Decentralized Identifier working group three years earlier in 2019—would become an official web standard. DIDs are a type of identifier that enables a verifiable, decentralized digital identity. A verifiable credential (VC), as defined by the specification, can represent all of the same information as a physical credential. With digital signatures, verifiable credentials are more tamper-evident and more trustworthy than their physical counterparts.

This approval will energize developer communities to create better and healthier standards for the internet. But it is equally important to acknowledge what DID v1.0 does **not** specify, and the work that remains. The DID v1.0 specification standardizes the DID syntax, a common data model, serialized representations, and operations. **DID v1.0 does not standardize methods of persistence, resolvability, decentralization, and cryptographic verification—leaving DID *methods* *under*-standardized.**

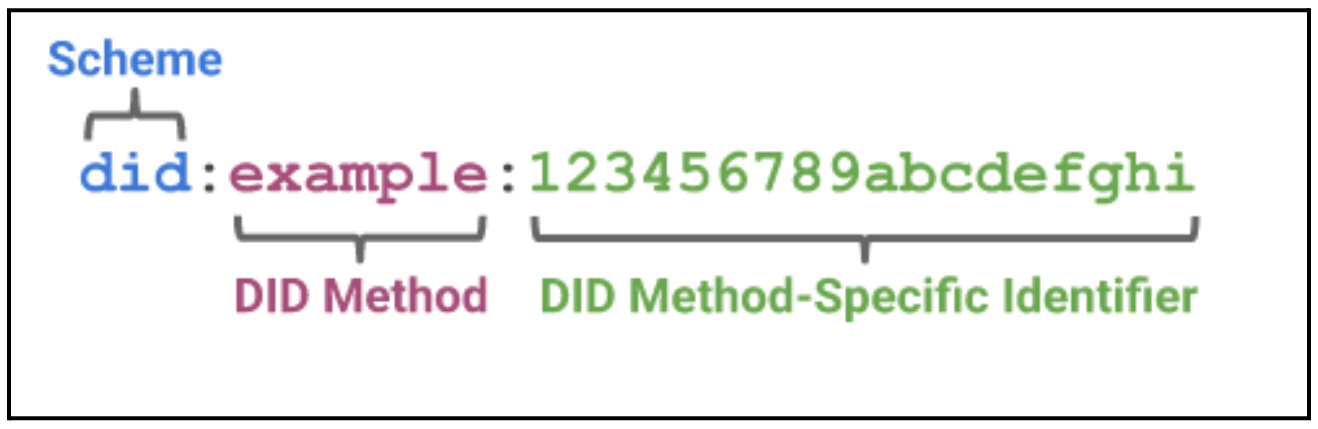

**Image 3.** Highlights the three-fold structure of DIDs.

DID methods are unique implementations of the normative DID v1.0 specification. By way of analogy, you can think of the DID v1.0 as a template for what a DID method *should* look like and the properties it *should* uphold, outlining how a globally unambiguous decentralized identifier should work. But each DID method differs in tactical implementation details on *how* to achieve that, implementing quasi-independent identifiers. Moreover, each DID method differs significantly in terms of technical frameworks and developer communities. Over 100 DID methods exist today, without consensus around which will crystallize into a standard. These DIDs also differ in type: ledger-based, ledger-middleware, peer, or static. Many of today's applications are only composable with a subset of these DID methods. Hence, Mozilla's objection to W3C:

> *The DID architectural approach appears to encourage divergence rather than convergence & interoperability. The presence of 50+ entries in the registry, without any actual interoperability, seems to imply that there are greater incentives to introduce a new method, than to attempt to interoperate with any one of a number of growing existing methods.*

Thus, the director encourages continued work "and search for consensus on standard DID methods." In fact, in the Web3 community did:key and did:pkh are generative DID methods that have minimal specifications and that are getting adoption across the web3 space already.

## Decentralized Naming Services & SBTs compared to DIDs & VCs

The design goals of DIDs, as noted in the specification, align with many web3 design goals: decentralization, privacy, security, proof-based, discoverability, interoperability, portability, simplicity, and extensibility. Although there are many blockchain-based DID methods, we use decentralized naming services as a starting point for discussion.

DIDs can be created and used in different ways, depending on their method. DIDs are considered to be globally unique as they are a type of uniform resource identifier (URI) without a centralized registration authority (e.g., no equivalent of ICANN for DNS). Verifiable credentials are signed attestations. An issuer can make a claim about a subject (identified by their DID or other URI) and for this credential to be considered a verified presentation, the claim needs to be signed by the subject.

DIDs share many properties with commonly used, distributed, and open naming services like Ethereum Naming Service (ENS). Though built on Ethereum, ENS can be connected to non-Ethereum machine-readable identifiers. When a DID is resolved, the DID document is dynamically built, exposing the minimal amount of information (public keys, service endpoints, and authentication methods) needed for downstream applications. Similarly, when an ENS identifier is resolved, the externally owned addresses—what is commonly referred to as a "wallet address"—or smart contract addresses are exposed. Additionally, once you have the public key after resolving the DID or ENS, individuals are capable of issuing VCs or SBTs respectively to the identifiers, making both paradigms vulnerable to negative reputation.

| | **Decentralized Identifiers (DIDs)** | **Ethereum Naming Service (ENS)** |
|---|---|---|
| Uniqueness across a defined domain | ✅ | ✅ |
| Persistent | ✅ | ✅ |
| Machine-friendly identifiers (strings based on random numbers / cryptography) | ✅ | ✅ |
| Resolvable | Defined by DID method | Uses ENS resolver function |
| Associated data is expressed | Located in DID documents | Located in public address transaction history. |
| Fully decentralized namespaces without delegation | ✅ | ✅[3] |
| Cryptographically-verifiable | ✅ | ✅ |
| User Control | Fully under control of the DID controller | Fully under control of the ENS controller |

**Table 1.** Illustrates differences between DIDs and ENS.[4]

[3] Domain name system (DNS) on the other hand has a hierarchical, delegateable namespace based on centralized root registries for top-level domain (TLD) names

[4] The DID information from this table was partly taken from the 2020 SSI Meetup Decentralized Identifier (DIDs) fundamentals deep dive.

## Negative Reputation

The "Decentralized Society: Finding Web3's Soul" paper sketched how non-transferrable NFTs, or "soulbound" tokens, could be a social identity primitive in web3. Its publication sparked debates about consent and the question of whether digital identities could become unwillingly and irrevocably tagged with negative information. For example, here is a quote from the Bankless interview titled "Soulbound" by the founder of Disco.xyz, Evin McMullen:

> "If Vitalik makes a claim that I am a total chad through a soulbound token ... apps will be able to read and access this information forever. Vitalik did not secure my consent to give me a token, which is why someone was able to send him billions of Shiba tokens.
>
> Allowing for consent in the soulbound token paradigm is a norm but is not enforceable even if it is possible to secure consent first. Since gaining consent will be seen as inconvenient, it will be routed around."

Decentralized naming services like ENS are vulnerable to non-consensual "scarlet letters" issued or dropped as a simple non-transferable NFT the user did not mint. A 'scarlet letter' as depicted in the novel is the forced display, and disclosure of a negative sentiment about oneself to others at all times against their will. In the digital context, this definition remains the same. But there are two important caveats to consider.

The first caveat is that non-transferable NFTs are the most primitive form of SBTs—which today lack a specification. In particular, the singular property of non-transferable NFTs is non-transferability, and they lack any of the nuanced socially programmable privacy or community recovery aspirations sketched in the DeSoc paper— pathways which theoretically may enable community mechanisms to counter certain forms of malicious tagging. Without these mechanisms, today's architecture of permissionless wallets allows for blockchain-based identity certificates to be sent to any wallet address (externally owned account) without the consent of the recipient. The contents of this certificate are potentially public, immutable, as well as challenging and expensive to revoke. It's important we consider finding mechanisms to address this issue.

The second caveat is that it would be a mistake to think that DIDs and VCs are inherently immune to scarlet letters, with all DID methods and associated VCs somehow protecting users against such non-consensual tags. The key difference between unsigned off-chain VCs and on-chain non-transferable NFTs whose subjects have not countersigned or minted them is where the data resides. In crypto wallets, scarlet letters are airdropped to a user's wallet without their consent. With DIDs, the scarlet letter lives on the data stream of the issuer and will only live on the subject's datastream should they countersign the claim.

To substantiate this point, we issue an innocuous "scarlet letter" to an account that uses the DID:3 method—a method created by the Ceramic network and used by various social apps today, including Disco.xyz and Orbis. We use Cerscan, a tool that indexes all streams on the Ceramic network, for this analysis.

The first step was to find a stream of data to which we could issue a credential. As such, our lead author found a user who has created a stream of data connected to their account on the Disco.xyz app. Every certificate, in the form of a VC, is a data stream originated by an issuer and pinned on IPFS via Ceramic nodes. The person's linked information is visible in the open, such as their Ethereum address, Twitter handle, and their `DID`.

```
"credentialSubject": {
  "id": "did:3:kjzl6cwe1jw145xum3kzylvgqqlut4u1zrzyhdmza8gf7cmvlr7ana5qh7kxm9j"

}
```

**Image 4**. Specific identifier for the DID:3 method found on the person's data stream.

Our author issued a credential to the person with his DID being the `credentialSubject`. As you can see in the credential that reads the content "*Disco and Souls are siblings!!*", and anyone can read and verify who issued the credential. That credential is an innocuous scarlet letter tagged to the person's DID from the issuer's data stream (confirmed by the credentialSubject id matching the DID id). See Image 5.

Retracting the claim would require the issuer to delete it from their node. But this does not necessarily solve the problem. Every Ceramic node has permission to read from any other Ceramic node, making it possible for another node—or even person—to copy and store data from a node, even if the data is later deleted. Thus, identity holders cannot ensure that scarlet letters get removed from the network or history of the internet entirely.

```
"claim": {
  "object": "Disco and Souls are siblings!!"
   },
"credentialSubject": {
   "id": "did:3:kjzl6cwe1jw145xum3kzylvgqqlut4u1zrzyhdmza8gf7cmvlr7ana5qh7kxm9j"
   }
  }
```

**Image 5.** Highlights both the claim made and `DID` of whom we issued it was issued that matches from the data stream in Image 4. You can find this claim and proof of the existence of the credential subject, here.

The DID specification states that “it is *critical* that DID documents contain no personal data.” This example and other exploratory projects illustrate that implementations today may be inconsistent with the specification, as the data stream identifies the twitter handle and public ETH address name, making it susceptible to scarlet letters. This underscores that current implementations are still experimental, and require substantial R&D and careful engineering before they become complete, generalizable, secure, or consistent with the standard. Thus, promotion as a standard may be a lower priority than iteration on it.

Certainly, these implementations are not a final product (as is the case with simplistic non-transferable NFTs). Many of these projects are still in an experimental phase—being in private-beta, beta, or on test-net (as is the case with Disco.xyz). For these projects, an identity holder's ability to curate—namely remove—tagged non-consensual verifiable credentials is the crux of their enthusiasm for these emerging primitives and they intend to embrace solutions commented later and not necessarily rely on Ceramic as the data layer they are built on top of.

So how could one stop scarlet letters on DIDs?

Section 3.3 of the verifiable credentials specification comments on verifiable presentations, which specify the components needed to present a VC in a verified way. One key component is the Presentation Proof Graph which requires the signature of the VC’s owner to be considered “verified.” However, as shown in Table 1., this does not stop scarlet letters as **one can still index widely-accessible online VCs.**

| **Outcomes** | **Side of Transaction** | **Soulbound Tokens (SBT)** | **Verifiable Credentials (VC)** |
|---|---|---|---|
| No consent. Publicly viewable. Indexable. | Sender | I issue an SBT to a public address. | I issue a VC to a public address. |
| Consent. Publicly Viewable. Indexable. | Receiver | I sign a transaction approving an SBT that I am eligible to claim. | I sign a VC that was issued to my public address. |

**Table 2.** Shows why the ‘scarlet letters’ are, in practice, the same for both SBTs and VCs.

In the VC specification, section 7.3 comments specifically on identifier-based correlation, which is a possible solution by preventing correlation of identities with one another and hiding the identifier during verifiable presentation. The specification reads:

> “disclosing the credential identifier (credential.id) leads to situations where multiple verifiers, or an issuer and a verifier, can collude to correlate the holder. If holders want to reduce correlation, they should use VC schemes that allow hiding the identifier during verifiable presentation.”

But hiding the identifier during a verification presentation does not thwart scarlet letters if the DID address is disclosed, as many users unwittingly do when signing up for such platforms, making themselves vulnerable to scarlet letters.

Moreover, anytime a DID interacts with another DID or application (whether it be on-chain or off-chain), it reveals itself. As noted in the Ceramic documentation, for a transaction to be successfully processed, the recipient’s DID document must be resolved, which means that a public key corresponding to the transaction signature must be present in the recipient’s DID document. So any interaction between a user and another recipient DID or even application reveals the user’s public address to the counterparty. With knowledge of the user's public address, anyone (whether it be another application or individual) can make a claim about the DID. Although these claims don’t sit on the user’s data stream (but the issuer’s who issue the unsigned claim to the DID), they can be easily indexed on in the network and mapped back to the initial user (as is possible with the internet today).

Storage of a scarlet letter on a different data stream doesn’t guard against negative reputation, but merely adds a step in discovering it. In particular, an API could scrape up claims about a DID across data streams, akin to search engines today making visible all scarlet letters. In fact, so long as people can freely publish content, someone can build a search engine to gather and store attestations, even if they are spread across data streams not controlled by the user. When it is in the application’s interest to be aware of negative reputation, conducting this search across the network will become part of the standard design solution[5]. In this sense, scarlet letters are both possible and inevitable with both substrates, as shown in Image 7, and reliance on data streams to protect against negative reputation is overdrawn.

[5] A user's data has the ability to transport with them in Web3. The benefits of network effects from Web2 companies no longer become a moat for success. As Co-Founder of Ceramic noted in a recent podcast , Web3 companies no longer need to be ten times better than Web2 incumbents to succeed, but rather serve a small community extremely well, now that the data is interoperable. This makes it even more compelling for Web3 teams to take the additional effort to perform this extra work on doing a search on the data layer about its smaller set of users that traditional Web2 companies may be deterred from doing.

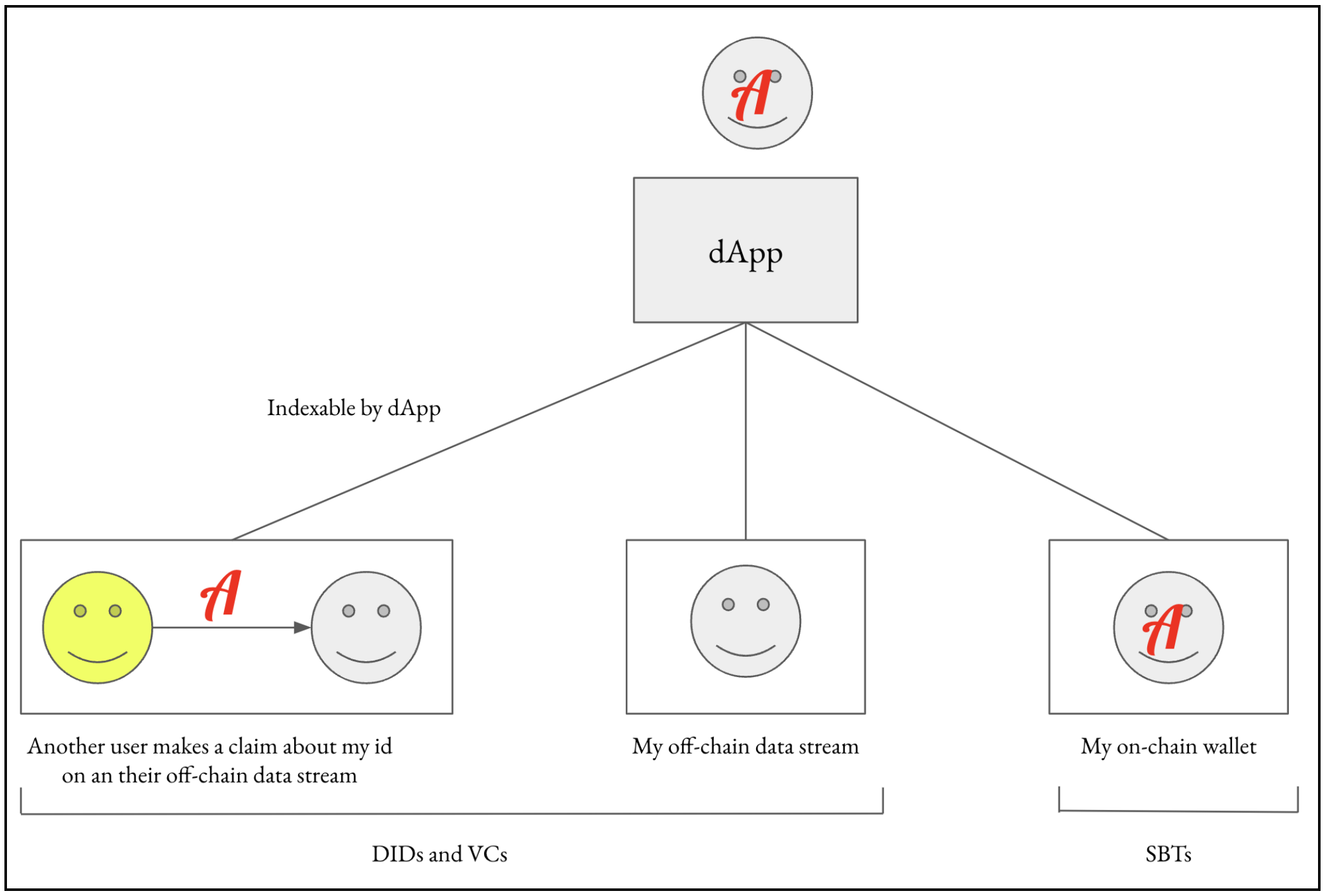

**Image 6.** Location of scarlet letters

The current lack of a "silver bullet" solution means we should keep an open mind about technological substrates and continue experimentation across different methodologies. The W3C director writes, “The DID core specification as it is now does not lack proofs of implementability. The histories of many Internet standards show that future work can — and often does — lead to improvements to an initial standard.” The translation from specification to scalable adoption for DID methods and VCs is nowhere settled, and will benefit from plural designs.

## Composability with Blockchain Code

So if unsigned VCs are in some ways conceptually equivalent to indexable SBTs whose subjects haven’t countersigned or minted them, what are the real tradeoffs between these two substrates? **This may be the most generative and productive debate to have today about decentralized identity substrates.** Rather than having the definitive answer, our aim here is to kickstart the conversation, by making the case that composability with blockchain code is one of them.

Composability is vital for innovation around communities like DAOs that use blockchains and blockchain-based certificates to coordinate and govern assets. Composability of identities with blockchains enables:

- **Correct execution of code**: "the blockchain accepts inputs ("transactions") from [identities], correctly processes them according to some predefined rules, and returns the correct output (or adjusts the blockchain's "state" in the correct way").
- **Censorship-resistance:** "any user that *wants* to send a transaction, and is willing to pay a high enough fee, *can* send the transaction and expect to see it quickly included on-chain."
- **Private** participation of identities in applications: "you should not be able to tell which candidate someone specifically voted for, or even if they voted at all."
- **Coercion-resistant** participation of identities in applications: "you should not be able to *prove* to someone else how you voted, *even if you want to."*[6]

To illustrate, imagine a community that operates an election through an Ethereum Virtual Machine (EVM). Suppose community members need a special membership certificate to participate in the election. If we store the membership certificates in an EVM compatible system as blockchain-based certificates (making it accessible at runtime), the EVM can execute the election function based on the membership certificates and vote inputs as programmed—hence, achieving the property of *correct execution*. Assuming the current distribution of membership certificates conforms to the state of this EVM, any voter authorized by the certificate can participate in the election with *censorship resistance*.

Privacy and coercion resistance are also rich areas of research and development in blockchain contexts. There are several implementations of zero-knowledge proofs under refinement that offer the promise of *privacy*: voting without revealing the content of a vote, or that a vote was ever made. Additionally, minimum anti-collusion infrastructure (MACI)—frameworks where voters are unable to disclose their choices are all well underway in research—thus offering the promise of future *coercion resistance.*

With the advent of these cryptographic tools and strong developer communities around them, **blockchain-certificates like SBTs have the potential to achieve the properties of privacy, correct execution, coercion-resistance and censorship resistance to the voting context.**

Web3 applications that utilize off-chain DIDs (for example on the Ceramic Network) are indeed compatible with most blockchains and addresses, but present important tradeoffs: mainly, any EVM-native (or blockchain-native) application must rely on proprietary off-chain databases or oracles[7] as a trusted intermediary to present the verifiable credentials as inputs. Consider the election example again. In the DID scenario, membership certificates would be stored as VCs mapped to off-chain DIDs. Community administrators and voters would have to rely on an oracle as a trusted intermediary to correctly execute the election function and recognize these VCs. In particular, if VCs are stored in an off-chain proprietary datastore, an oracle might strategically choose to blacklist a VC, introducing censorship.

---

[6] In Vitalik Buterin's (2021) blog post, you can learn more about these concepts applied to the blockchain voting context: "Blockchain voting is overrated among uninformed people but underrated among informed people."

[7] Beniiche's "A Study of Blockchain Oracles" defines an "oracle as an interface that delivers data from external data sources outside the blockchain for a smart contract to consume". Oracles can differ in their origin (software, hardware, or human), direction of information (inbound or outbound), and in their trust guarantees (centralized v. decentralized).

Moreover, off-chain solutions make resolving disputes hard. Blockchain oracles today are most often used for data sources that are widely available, uncontested and not subjective (e.g., price of ETH in USD). Off-chain identity data, however, is idiosyncratic. Dispute resolution methods will vary. It's unclear both how to surface disputes *before* the oracle yields outcomes on-chain, and more challenging how to resolve them *after*. Data that is cryptographically verifiable and stored off-chain (like Ceramic), could in theory be verified on-chain and be made possible at scale with ZKPs in the future.

There are several ways off-chain solutions might achieve the properties desired for composability. For example, the problem of correct execution could be solved by implementing a verifier for a VC scheme in solidity or vyper. Additionally, if these VCs are stored off-chain in a publicly accessible datastore (e.g., IPFS), then it could become impossible for an oracle to censor accounts without negative repercussions like slashing. Moreover, there has been work to enable minimal disclosure with DIDs and VCs with BBS+ signatures and CL Signatures to improve privacy for settings like the election example discussed. It is also reasonable to assume that MACI frameworks discussed could be adapted to off-chain settings to enable coercion resistance.

Additionally, it's important to recognize that there is ongoing research and work in progress to advance the integration of VCs and blockchain code. Work on public key infrastructure (PKI), for example, is particularly promising. Today, Gitcoin Passport holders manage their Ceramic DIDs via the did:pkh method. So Passport holders can use their Ethereum keys to control Ceramic streams. This doesn't solve the oracle problem, where Ceramic-based identity certificates rely on a trusted intermediary to input the VCs onchain. But it is one step toward composability, improving the experience of Passport holders and developers.

Advocates of DIDs and VCs often tout the data portability advantages that DIDs have compared to SBTs. Yet, the analysis above should reveal that DIDs and VCs require further research and development before they can interoperate with blockchain apps or larger blockchain platforms without an oracle without attendant trade offs in correct execution and censorship resistance. Moreover, there are other significant challenges with recovering encryption keys and indexing (search on off-chain data) which make DIDs and VCs challenging for web3 communities. As such, **it's still unsettled if web3 can benefit from VCs and DIDs more than from blockchain-native identity (ie., SBTs).**

## Identity & Smart Contract Composability Today

Let's walk through options for making identity data composable with smart contracts today.

Consider a DAO for the alumni of the University of Washington. The DAO's goal is to aggregate funds from alumni and invest them into alumni causes. This DAO requires you to prove that you are an alumnus of the University of Washington. So you need some way of showing the smart contract that you are an alumnus. How can we do this?

1. Put this data into an ERC-721 token issued by the University of Washington. This solution is suboptimal as it does not preserve any privacy, is costly at scale, and the permit would be transferable to others at a low cost.
2. Call the smart contract with (if on Ethereum) the EIP-712 method. This lets you put a JSON-LD-based credential as the payload into the blockchain code call. Gas concerns aside, this would work if Ethereum did not emit these logs. Anyone that can run an archive node on Ethereum could rerun the chain and see what the payload was for all time. This is also suboptimal as it does not preserve privacy.
3. You could (as mentioned in the DeSoc paper) have a pointer to a VC that sits off-chain from an on-chain ERC721/ERC1155 token. This is useful for controlling and interacting with a front-end. Still, the blockchain apps themselves cannot read the data that sits off-chain.
4. You could prove to a front-end (whoever controls the smart contract) that you have a VC with some selective disclosure mechanism. This front-end has the keys to write to the smart contract and put you on an "allow-list." This method would avoid putting PII on-chain but contextually leaks privacy since you have met some conditions to put you on the allow list.
5. Once it matures, you could use a version of our zero-knowledge soulbound token implementation that Enrico Bottazzi and this article's lead author recently released in open-source. This mechanism is not mature, not very scalable, and has many computational limits with most browsers today.

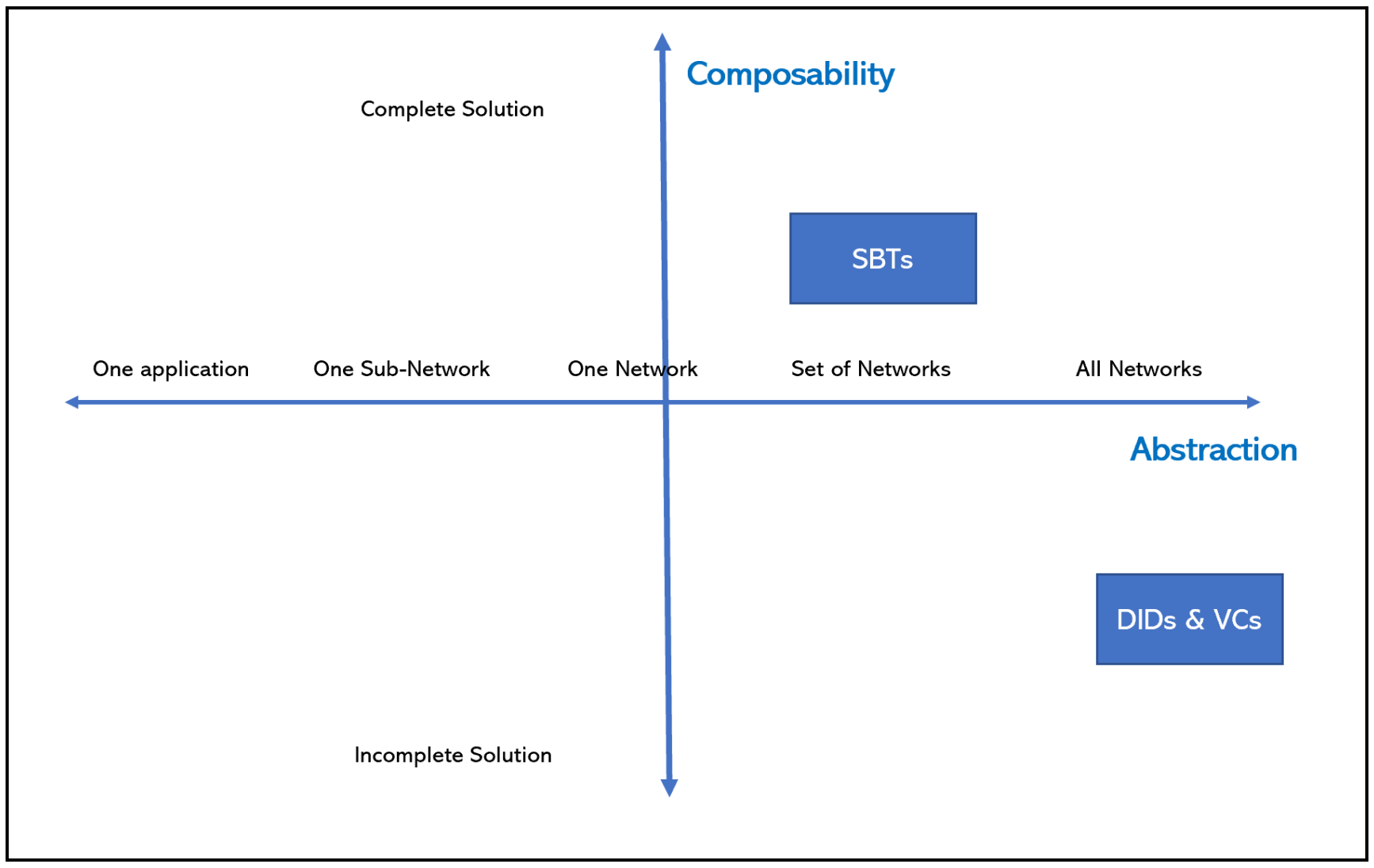

**Image 7.** This shows where current identity solutions sit in this matrix.

## Moving Forward

Pluralism in technology recognizes that technology is fundamentally assistive; technology should serve the needs of communities—rather than the community serving the needs of technology. We should be cautious with the pace that we adopt implementations of standards, and impose them on other communities when they are underdeveloped and contested, and moreover present significant tradeoffs to the community's goals.

Technology is also dynamic. Networks evolve. Innovation is a process, not an endpoint. We should generally be skeptical of one-size fits all solutions, or "final solutions," and instead embrace plural solutions. Rather than framing DIDs/VCs and SBTs as either/or solutions, we consider them as i*nnovation complements* with different tradeoffs for different communities who know their needs best. We encourage continued collaboration in a dialectic where the synthesis can become greater than either paradigm alone. We are still early.